\RequirePackage{lineno}
\documentclass[aps,prl,reprint,preprintnumbers,nobibnotes,nofootinbib,floatfix]{revtex4-1}

\usepackage[utf8]{inputenc}
\usepackage[english]{babel}
\usepackage{blindtext}
\usepackage{amssymb,amsmath}
\usepackage{bbm,latexsym}
\usepackage{slashed}
\usepackage{cancel}
\usepackage{xcolor}

\usepackage{graphicx}
\makeatletter

\providecommand{\tabularnewline}{\\}

\usepackage[]{subcaption}

\usepackage[plainpages=false,pdfborder={0 0 0},linkbordercolor={1. 1. 1.},citebordercolor={1. 1. 1.},urlbordercolor={1. 1. 1.}]{hyperref}

\newcommand{\eg}{\textit{e.g.}\ }
\newcommand{\ie}{\textit{i.e.}\ }

\newcommand{\msbar}{$\overline{\text{MS}}$}
\renewcommand{\eqref}[1]{Eq.~(\ref{#1})}

\begin{document}
\title{Vacuum expectation value renormalization in the Standard Model and beyond}

\author{Vytautas Dūdėnas}
\email{vytautasdudenas@inbox.lt}
\affiliation{Institute of Theoretical Physics and Astronomy, Faculty of Physics, Vilnius University,	9 Saulėtekio, LT-10222 Vilnius, Lithuania}

\author{Maximilian Löschner}
\email{maximilian.loeschner@kit.edu}
\affiliation{Institute for Theoretical Physics, Karlsruhe Institute for Technology,  Wolfgang-Gaede-Straße 1, 76131 Karlsruhe, Germany}

\preprint{KA-TP-18-2020, P3H-20-063}

\begin{abstract}
We show how the renormalization constant of the Higgs vacuum expectation value, fixed by a tadpole condition, is responsible for gauge dependencies in various definitions of parameters in 
the $R_{\xi}$-gauge.
Then we show the relationship of this renormalization constant to the Fleischer-Jegerlehner (FJ) scheme, which is used to avoid these gauge dependencies. 
In this way, we also present a viewpoint on the FJ-scheme complementary to the ones already existing in the literature.
Additionally, we compare and discuss different approaches to the renormalization of tadpoles by identifying the similarities and relations between them.
The relationship to the Higgs background field renormalization is also discussed. 
\end{abstract}

\maketitle

\section{Introduction}

In modern particle physics, high precision calculations are of increasing importance for finding signs of new physics in the comparisons of theory predictions to experimental data.
An integral part of these calculations is the subject of renormalization.
Even though the main principles of renormalization are well understood  (see \eg \cite{Denner:2019vbn} for a recent review of electroweak radiative corrections) and represent a standard textbook subject, some subtleties remain being actively discussed.
One of these is the subject of vacuum expectation value (VEV) renormalization in conjunction with so called tadpole schemes.
For existing examples of discussions in the literature, see \eg \citep{Grassi:2001bz,Weinberg:1973ua,Actis:2006ra,Krause:2016oke,Dudenas:2018wlr,Grimus:2018rte,Denner:2016etu} or \citep{Denner:2019vbn} for a list of tadpole schemes.
However, we find that a unified exposition of the
relationships between such schemes is still missing in the literature.
Hence, in this paper, we want to elucidate the relation between the renormalization of vacuum expectation values, tadpole schemes, gauge dependencies and the special role of Goldstone boson tadpoles in this respect.
More specifically, we show the connections between methods that are commonly used in precision calculations for the Standard Model (SM) as e.g. in~\citep{Denner:2019vbn} and more formal discussions of VEV-renormalization in general gauge theories as \eg in~\citep{Sperling:2013eva,Sperling:2013xqa}.

We want to emphasize the known fact that an independent VEV (or tadpole) renormalization
constant is necessary in addition to the renormalization of the parameters
and fields of the unbroken theory in order to render all $n$-point Green's functions finite
in  $R_\xi$-gauge 
(this was already noted in \eg \cite{PhysRevD.8.1747,Chankowski:1991md}).
In the broken phase, the usage of the VEV in gauge-fixing
functions affects the global symmetry properties of the theory \citep{Sperling:2013eva},
which leads to the need of this additional degree of freedom\footnote{
	Note that in some older literature, when the $R_\xi$-gauge was not as commonly used as it is nowadays, this fact might not be mentioned. As an example, in~\cite{Lee:1972fj,Lee:1974zg,Lee:1972yfa}, this fact is not discussed due to the use of $R$-gauge fixing.}. Hence
in spontaneously broken gauge theories such as the SM, this can be
understood as an artifact of the gauge-fixing procedure rather than
a direct consequence of the mechanism of spontaneous symmetry breaking
itself. This introduction affects definitions of parameters, leading to gauge dependencies in some of them. 
In principle, these gauge dependencies will always cancel in physical observables.
Moreover, the S-matrix can even be made finite without renormalizing tadpoles at all, hence, leaving the one-point Green's functions infinite \cite{PhysRevD.8.1747}.
Nevertheless, it can be favorable to demand gauge-independent physical parameter definitions in perturbative calculations, \ie for intermediate expressions such as mass-counterterms.
The Fleischer-Jegerlehner tadpole scheme (FJ-scheme) \citep{Fleischer:1980ub} was proposed to avoid these spurious gauge dependencies.
This becomes even more important if one goes beyond the Standard Model (BSM), where the usual on-shell (OS) renormalization is not possible for all parameters (see \cite{Freitas:2002pe} for a discussion of the problems arising here) 
and explains part of the renewed attention
to the subject in the context of the two Higgs doublet models \citep{Grimus:2018rte,Denner:2016etu,Dudenas:2018wlr,Krause:2016gkg,Krause:2016oke}. 

The FJ-scheme is closely related to the aforementioned additional
independent VEV-renormalization constant. 
As we will see, the FJ-scheme
makes sure that this degree of freedom would not enter the parameter and
counterterm definitions and hence, allows for gauge-independent definitions. 
However, neither in the original paper, 
nor in the more recent ones on the scheme, this relationship is explicitly exposed. 
Instead, the notions  \emph{proper} VEV \cite{Fleischer:1980ub} or the \emph{correct one-loop minimum} \cite{Denner:2019vbn,Denner:2016etu,Krause:2016oke,Krause:2016gkg} are used to motivate the gauge cancellations in the FJ-scheme.
We try to fill the gap by exposing this feature and also suggests to look at the scheme as being simply a convenient set of counterterm redefinitions.
Using this viewpoint, we also show translations between different tadpoles schemes. 

With a pedagogical purpose in mind, we carry our study out for the SM at the one-loop level, but also comment on implications to BSM.
Moreover, the paper is set up such that our results can be easily reproduced
using \eg the native FeynArts \cite{Hahn:2000kx} SM-file together with FeynCalc \cite{MERTIG1991345, Shtabovenko_2016, Shtabovenko_2020} or FormCalc \cite{Hahn:1998yk,Hahn:2016ebn} (both with and without background fields).

Note that in order to extend our discussion of gauge-dependencies to higher loop-orders, one has to adopt the Complex Mass Scheme (CMS) \cite{Sirlin:1991fd,Denner:1999gp, Gambino:1999ai,Denner_2005,Denner:2006ic,Kniehl:2001ch,Espriu:2002xv} in addition to using the FJ-scheme.
The reason being that in the presence of unstable particles, propagator poles can acquire imaginary parts from two loops onward  and the usual OS-scheme leads to gauge dependent mass definitions in that case. 
One can therefore only prove the gauge-independence of the complex propagator poles \cite{Gambino:1999ai} and needs to include this in the discussion via the CMS.

We start our presentation in Sec.~\ref{sec:Additional ct}, by explaining why it is necessary to have an additional renormalization constant in the spontaneously broken phase of the SM as compared to the unbroken phase and introduce a tadpole condition to fix the former.
In Sec.~\ref{sec:translation of renconsts}, we present the translations between the renormalization constants of different parameter sets that are used as independent in the renormalization procedure. 
In particular, we show the relations between renormalization constants of symmetry-based (or "original") parameters of the theory, to the ones used in the usual OS-scheme in~\cite{Denner:1991kt}.
These translations illuminate the gauge dependencies in the definitions of the usual mass renormalization constants. 
In Sec.~\ref{sec:relating-schemes}, we define the FJ-scheme as known from the literature, show how it provides gauge-independent counterterm definitions and present a new viewpoint on the scheme in terms of renormalization constant reparameterization. 
This section also relates the FJ-scheme to the findings presented in~\cite{Sperling:2013eva} and moreover, we illustrate the differences between the tadpole schemes by comparing VEV-renormalization constants numerically and comment on the outcomes.
We conclude our presentation in Sec. \ref{sec:conclusions}.
Some details of our calculations can be found in the appendices. These include a short note on the construction of the $R_\xi$-gauge in the background field formalism in App.~\ref{app:Gauge-fixing-functions}; the calculation of the purely gauge-dependent divergences using the background fields in the SM (an adaptation from \cite{Sperling:2013eva}) in App.~\ref{app:Calculation-using-BRS}; consequences of different renormalization conditions of this approach in App~\ref{app:OS gauge dependence}; explicit divergences of renormalization constants in App.~\ref{app:explicit divs} and numerical input values that we used to calculate VEV-renormalization constants in App.~\ref{app:numerics}. 

\section{An additional counterterm in \texorpdfstring{$R_\xi$}{Rxi} } \label{sec:Additional ct}

The necessity of an independent VEV-renormalization constant for renormalizing one-point Green's functions has been noted before (\eg see \cite{PhysRevD.8.1747,Boehm:1333727,Chankowski:1991md}).
However, one might come to the conclusion that this is done purely for convenience.
Reasons for this impression are given by the fact that S-matrix elements are finite even without renormalizing tadpoles at all \cite{PhysRevD.8.1747} or the fact that a gauge-fixing other than the $R_\xi$-gauge is used, such as in~\cite{Lee:1972fj,Lee:1974zg,Lee:1972yfa}, where indeed all $n$-point Green's functions can be made finite via multiplicative renormalization of the parameters of the unbroken theory.
Here, we want to clarify that the latter statement is not true in $R_\xi$-gauge.
As discussed rather recently in~\cite{Sperling:2013eva}, the explanation comes from the fact that this gauge-fixing explicitly breaks a global $SU(2) \times U(1)$ symmetry.
Then, a VEV-counterterm can not be forbidden on the grounds of symmetry arguments or in other words, its divergence structure is not fixed by the field strength renormalization of a physical scalar.
In order to fix its divergence structure, the authors of \cite{Sperling:2013eva} restore the original global symmetries of the theory by the introduction of background fields\footnote{This set-up is also related to a more formal studies of algebraic renormalization \cite{Piguet:1995er,Haussling:1996rq,Kraus:1997bi,Kraus_1995,Kraus:1998xt,Kraus:1998ud} . 
	In these studies, the term \emph{rigid symmetry} is used instead of \emph{global symmetry}, but the meaning is the same.}
and express the VEV renormalization in terms of the background field renormalization.
Additionally, via the use of Becchi-Rouet-Stora-Tyutin (BRST)-sources, the difference between the VEVs divergence structure and the physical scalars field renormalization is isolated.
We will now clarify these points explicitly for the SM.

We consider the usual Higgs potential of the SM:
\begin{equation}
V\left(\phi\right)=\mu^{2}\phi^{\dagger}\phi+\lambda\left(\phi^{\dagger}\phi\right)^{2}\,.\label{eq:higgs potential}
\end{equation}
The neutral component of the Higgs doublet $\phi$ acquires a VEV $v$ as in
\begin{equation}\label{eq:higgs doublet}
\phi=\left(\begin{array}{c}
G_{W}^{+}\\
\frac{1}{\sqrt{2}}\left(v+h+iG_{Z}\right)
\end{array}\right),
\end{equation}
where $G_W^+$ and $G_Z$ are the Goldstone boson fields and $h$ is the physical Higgs field.
The tree-level minimum condition for \eqref{eq:higgs potential} gives
\begin{equation}\label{eq:minimum}
\frac{\partial V}{\partial h}|_{h=G_{W/Z}=0}=0\Rightarrow\,v^{2}=-\frac{\mu^{2}}{\lambda},
\end{equation}
which leads to the Higgs mass being:
\begin{equation}
m^2_h = 2\lambda v^{2}=-2\mu^{2}.
\end{equation}

Following \cite{Sperling:2013eva}, we introduce background fields denoted by ``hats" via
\begin{equation}
\phi\to\phi+\hat{\phi}=\left(\begin{array}{c}
G_{W}\\
\frac{1}{\sqrt{2}}\left(h+iG_{Z}\right)
\end{array}\right)+\left(\begin{array}{c}
\hat{G}_{W}^{+}\\
\frac{1}{\sqrt{2}}\left(v+\hat{h}+i\hat{G}_{Z}\right)
\end{array}\right).\label{eq:quant+back}
\end{equation}
The $R_{\xi}$-gauge-fixing function is modified in such a way that the gauge-fixing and
ghost part of the Lagrangian are invariant under the global gauge transformation, where the gauge parameters are restricted to\footnote{An explicit calculation shows that at one-loop, it is enough to use a single additional counterterm even
without the equality of the gauge parameters of \eqref{eq:gauge param equal}. However, it is unclear whether this holds at higher loop orders.
One can construct a gauge-fixing function which preserves the global invariance even when \eqref{eq:gauge param equal} 
does not hold instead.
Such a gauge-fixing function was introduced in~\cite{Kraus:1998ud}, 
yet we want to focus our discussion on the widely used $R_\xi$-gauge.
}
\begin{equation}
\xi=\xi_W=\xi_Z=\xi_A\,. \label{eq:gauge param equal}
\end{equation}
An explicit construction of the gauge-fixing functions using background fields in the SM is explained in App.~\ref{app:Gauge-fixing-functions}.
In this case, one finds that all $n$-point Green's functions can be rendered finite using only multiplicative renormalization constants,
\begin{equation}
	p \rightarrow Z_p p, \quad f\rightarrow \sqrt{Z_f} f, \quad \hat{f}\rightarrow \sqrt{Z_{\hat{f}}} \hat{f}, \label{eq:renormalization constants}
\end{equation}
for all parameters $p$, fields $f$ and background fields $\hat{f}$ of the theory.
However, in order to isolate purely gauge dependent divergences\footnote{Note that $\bar{Z}_{\phi}$ can nevertheless have gauge-dependent finite parts. 
This effectively comes about due to \eg the Passarino-Veltmann function $B_0(p^2, m^2 \xi, m^2\xi)$ carrying gauge-independent UV divergent, but gauge-dependent finite terms.
Hence, $\bar{Z}_{\phi}$  can just be used  as tool to disentangle divergence structures.
}
in a single constant $\hat{Z}_\phi$,
we write the field and background field renormalization as
\begin{equation}
\phi+\hat{\phi}\to\sqrt{\bar{Z}_{\phi}}\left(\frac{1}{\sqrt{\hat{Z}_{\phi}}}\phi+\sqrt{\hat{Z}_{\phi}}\hat{\phi}\right)\, ,\label{eq:renormalization+bf}
\end{equation}
and identify the renormalization of the components of the physical scalar and VEV by 
\begin{subequations}\label{eq:renorm-fields-vev}
\begin{align}
\sqrt{Z_h} &= \sqrt{\bar{Z}_{\phi} / \hat{Z}_{\phi}}\,, \quad \label{eq:field ren} \\ 
Z_{v} &=\sqrt{Z_{\hat{h}}}=\sqrt{\bar{Z}_{\phi}\hat{Z}_{\phi}}\,.\label{eq:vev=00003Dback}
\end{align}
\end{subequations}

It is now comfortable to introduce, schematically,
\begin{equation}
\delta = Z-1, 
\end{equation}
for every renormalization constant.
Then from \eqref{eq:field ren} and \eqref{eq:vev=00003Dback} at one loop
we have:\footnote{Note that since we started with the multiplicative constants, all parameter renormalization constants including the one of the VEV are defined dimensionless, leading to simpler relations between the constants.
The translation to the dimensionful constants can be easily done by replacing
\begin{equation}
\delta_p \to \frac{\delta p }{ p} \label{eq:dimensionless to dimensionful}
\end{equation}
in all our expressions. }
\begin{equation}
\delta_h = \bar{\delta}_{\phi} - \hat{\delta}_\phi\,,\quad \quad \delta_v  = \frac{1}{2} (\bar{\delta}_{\phi} + \hat{\delta}_\phi ). \label{eq:field and vev}
\end{equation}
It is thus clear that the VEV renormalization can coincide with the Higgs field renormalization only if $\hat{\delta}_\phi$ vanishes.
We show how one can get the divergent part of $\hat{\delta}_\phi$ 
 with the help of BRST-sources in App.~\ref{app:Calculation-using-BRS}.
The result for the UV-divergent part is
\begin{align}
	\hat{\delta}_{\phi}|_{\text{UV}} =  \frac{2}{4-D} \, \frac{\xi}{16 \pi^2 v^{2}}&\left( 2 m_{W}^{2} + m_{Z}^{2} \right ),
	\label{eq:UV div}
\end{align}
where $D$ is the number of space-time dimensions, $m_W$ and $m_Z$ are masses of W and Z bosons respectively.
Hence, it clearly vanishes for $\xi \to 0$, and can \emph{only then} let the VEV-renormalization coincide with the Higgs field renormalization.

To further see the role of $\hat{\delta}_\phi$, we consider the Higgs one-point function. 
By inserting renormalization constants into \eqref{eq:higgs potential}, collecting all the terms linear in $h$, 
and using the tree-level minimum condition \eqref{eq:minimum} we get a counterterm for the one-point function of $h$, \ie 
\begin{equation}
	\delta t_{h}=-\lambda v^{3}\left(\delta_{\lambda}-\delta_{\mu^{2}}+\bar{\delta}_{\phi}+\hat{\delta}_{\phi}\right),\label{eq:tadpole counterterm}
\end{equation}
which is fixed by the \emph{tadpole condition}, \ie
\begin{equation}
	\delta t_{h}+T_{h}=0, \label{eq:full tadpole condition}
\end{equation}
where $T_h$ are the one-loop tadpole contributions.
Throughout the rest of this paper, we keep this as a fixed condition in all different tadpole schemes.
The gauge-dependent part of the tadpole function, $T_h$, is
	\begin{align} \label{eq:Tadpoles-with-Goldstone}
		T_h^{(\xi)} &=
		\begin{gathered}
			\includegraphics{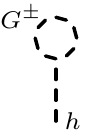}
		\end{gathered}
		+
		\begin{gathered}
			\includegraphics{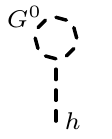}
		\end{gathered}
		\nonumber \\
		& = 
		\frac{1}{16 \pi^2} \frac{2 \lambda v^2 }{v} \Big[ A_0 (\xi m_W^2) 
		+\frac{1}{2}A_0(\xi m_Z^2)\Big],
	\end{align}
where $(\xi)$ denotes that we take only the gauge dependent tadpoles and $A_0$ is a one-point Passarino-Veltman function \cite{Passarino:1978jh}.
Checking the UV divergences in \eqref{eq:Tadpoles-with-Goldstone} and using \eqref{eq:tadpole counterterm} in \eqref{eq:full tadpole condition}, we see that the gauge dependent divergences cancel as
\begin{equation}
	T_{h}^{(\xi)}|_{\text{UV}}-\lambda v^{3}\hat{\delta}_{\phi}|_{\text{UV}} =0, \label{eq:tadpole gauge condition}
\end{equation} 
hence $\hat{\delta}_\phi$ alone absorbs all the gauge dependent divergences in the tadpole condition.
As shown in~\cite{Grimus:2018rte} for the multi-Higgs Doublet SM, one can come to the same conclusions by simply demanding the finiteness of all scalar $n$-point functions via $\{\delta_{\mu^2},\delta_\lambda, \hat{\delta}_\phi, \delta_h\}$, where $\hat{\delta}_\phi$ is introduced \emph{ad hoc} as an additional VEV-renormalization constant and without any reference to the background field formalism\footnote{In \cite{Grimus:2018rte}, $\delta v_k$ is the equivalent of $ v \hat{\delta}_\phi$.}.
For a derivation of this kind in the SM, we show the divergences of the relevant $n$-point functions in App.~\ref{app:explicit divs}.  
  
When using the $R_\xi$ gauge-fixing in the SM,
there are no gauge-dependent divergences in other renormalization constants (see App. \ref{app:explicit divs} for explicit expressions), hence we have
\begin{equation}
	\frac{\partial}{\partial \xi}  \delta_{\lambda}|_{\text{UV}} = \frac{\partial}{\partial \xi} \delta_{\mu^{2}}|_{\text{UV}} =  \frac{\partial}{\partial \xi} \bar{\delta}_{\phi}|_{\text{UV}}  = 0 \,. \label{eq:gauge independence of constants}
\end{equation} 
Since $\hat{\delta}_{\phi}|_{\text{UV}}$ vanishes when $\xi\to0$,
while the gauge-independent part is obviously untouched by this limit,
we can conclude that $\hat{\delta}_{\phi}$ renormalizes
purely the spurious divergences that are caused the gauge-fixing procedure in $R_\xi$-gauge. 
It is therefore only necessary as an independent renormalization constant when $\xi \neq 0$.
In this sense, it is the minimal addition to the set of renormalization parameters of the unbroken theory in order to render all $n$-point Green's functions finite. 
Moreover, any inclusion of $\hat{\delta}_\phi$ in counterterm definitions will carry over its gauge dependence.
As we will see in the next section, this is usually the case for mass counterterms. 

\subsection*{Remarks}

Employing the tadpole condition of \eqref{eq:full tadpole condition} is of special interest
when working with the 1PI generating functional which is defined as the Legendre transformation of the connected generating functional.
This transformation is only well-defined for vanishing one-point Green's functions, making \eqref{eq:full tadpole condition} essential \cite{Gambino:1999ai, Becchi:1996dt, Itzykson:1980rh}.
The 1PI generating functional in turn is used for deriving functional identities such as Slavnov-Taylor \cite{Slavnov:1972fg} or Nielsen identities
\cite{Nielsen:1975fs}. 

Questions about gauge-dependence in connection to VEV-renormalization can also be very relevant in
studies of effective loop-potentials.
It has already been noted in~\cite{Alexander_2009}, in the context of an Abelian-Higgs model, that Goldstone boson tadpoles violate the so called Higgs-low-energy theorem (HLET).
This relates to an older finding that 
one can not get Goldstone boson tadpoles from any potential by taking a derivative with respect to the Higgs VEV~\cite{Weinberg:1973ua}. 
The violation of the HLET is in correspondence to necessity of $\hat{\delta}_\phi$ in $R_\xi$-gauge.
In \cite{Alexander_2009}, the $R_\xi$-gauge is traded for the so called $R_{\xi,\sigma}$-gauge which reinstates a global symmetry of the Lagrangian and in this way avoids the necessity of an independent VEV counterterm though.

Similarly, in studies of finite temperature phase transitions (see \eg \cite{Patel:2011th, Garny:2012cg}), one finds gauge-dependent positions of the minima of effective loop-potentials.\footnote{Nevertheless, the values of the potentials at these points are found to be gauge-independent.}
Here, this is a result of using an $R_\xi$-gauge and defining the effective action as the sum of 1PI-graphs, \ie without tadpole and other external leg contributions.
Moreover, it is interesting to find the diagram in Fig.~1 of \cite{Garny:2012cg}, which determines the Nielsen coefficient of a one-loop effective potential.
Our definition of $\hat{\delta}_\phi$ via BRST sources shown in Fig.~\ref{fig:BRS loop} of App.~\ref{app:Calculation-using-BRS} is the equivalent to this in terms of its divergence structure.

As a last note in this section, we would like to stress the differences
in the use of background fields in~\cite{Sperling:2013eva} and \cite{Denner:1994xt}.
It might appear as if the authors state exact opposites, namely that a non-zero VEV-counterterm is strictly 
necessary versus the statement that no VEV-renormalization in addition to the Higgs 
field renormalization is needed.\footnote{Both references use the notation $\delta v$ for 
different quantities leading to a potential confusion.}
However, both statements are not contradictory
as the respective contexts differ.
In \cite{Denner:1994xt}, the authors do not renormalize quantum fields at all as they are interested only in the Green's functions of the background fields.
Then, the statement that no genuine VEV-counterterm is needed translates to the fact that no renormalization in addition to \eqref{eq:vev=00003Dback} is necessary.
In \cite{Sperling:2013eva} however, the focus lies on the renormalization of quantum fields, while the background fields are still used to preserve the symmetry structure of the theory.
Then, $\hat{Z}_{\phi}$ in \eqref{eq:vev=00003Dback} is interpreted
as an additional counterterm to the one of \eqref{eq:field ren}, due to a mismatch  between quantum field renormalization and the renormalization of its VEV.
In their notation, $\delta v \neq 0$ as they parameterized it in a relationship with the quantum instead of the background field renormalization as compared to \cite{Denner:1994xt}.

\section{Translation of renormalization constants \label{sec:translation of renconsts}}

Before a renormalization procedure is carried out, one has to choose a set of independent renormalization constants.
In the SM, one usually chooses experimentally well accessible physical parameters as independent renormalization constants while in BSM studies, it can be convenient to use the set of ``original'' theory parameters and the VEVs, especially when the use of an \msbar-scheme can not be avoided.
For the comparability of different choices, it is instructive to have a translation between the sets of renormalization constants.
To get these relations, consider we have a 
parameter set ${\{p\}}$ related at tree-level to a parameter set $\{p^\prime\}$ by some function $f$:
\begin{equation}
p^\prime_i=f_i(\{p\})\,.
\end{equation}
Introducing renormalization constants as in \eqref{eq:renorm-fields-vev} and expanding to one-loop order induces the relations
\begin{equation}
\delta_{p^\prime_i} = \frac{1}{ f_i (\{ p \} ) } \cdot  (\delta_{p_j} p_j ) \frac{\partial}{\partial p_j } f(\{p\}). \label{eq: relations between constants}
\end{equation}

In the SM, the relevant set of tree-level relations is
\begin{align}\label{eq:SM-relations}
&m_h^2 = \mu^2 + 3 \lambda v^2 , \quad m_W = \frac{v}{2}g_2, \quad m_Z = \frac{v}{2} \sqrt{g_1^2 + g_2^2},\nonumber \\
& e = \frac{g_1 g_2}{\sqrt{g_1^2 + g_2^2}}, \quad t_h= - v(\mu^2 + v^2 \lambda).
\end{align}
Using \eqref{eq: relations between constants}, we get the relations between the renormalization constants of the Parameter Renormalized Tadpole Scheme (PRTS) \cite{Denner:1991kt}, usually used with an OS-scheme, to the ones of the ``original'' parameters, namely  
\begin{equation}\label{eq:CT-parameters-sets}
\{t_h, m_h, m_W, m_Z, e\} \leftrightarrow \{v, \mu^2, \lambda, g_1, g_2\} ,
\end{equation}
where $g_1$ and $g_2$ are the $U(1)$- and $SU(2)$-couplings, respectively. 
The results are shown in the first column of Table \ref{tab:Application-of-an}, 
where $\delta_v$ is expressed in terms of the field renormalization constants of \eqref{eq:field and vev}.
This is to show where the gauge-fixing induced $\xi$-dependencies appear via $\hat{\delta}_\phi$. 
Inspecting the first column of Table \ref{tab:Application-of-an}, one clearly sees that $\hat{
\delta}_\phi$ enters the definition of the usual mass counterterms.
This means that the latter are necessarily gauge-dependent if one defines them as in the PRTS.
In the next section, we will present the FJ-scheme and show its role in the cancellation of gauge dependencies by virtue of the renormalization constant redefinitions shown in the second column of Tab.~\ref{tab:Application-of-an}. 

\begin{widetext}
	
	\bgroup
	\def\arraystretch{2.5}%
	\begin{table}
		\begin{tabular}{|c|c|}
			\hline 
			Usual tadpole scheme \cite{Denner:1991kt}  & FJ-scheme \cite{Fleischer:1980ub} \tabularnewline
			\hline 
			$\Delta=0$ & $\Delta=  \frac{T_{h}}{v m_{h}^{2}}= \frac{1}{2}\left(\delta_{\lambda}-\delta_{\mu^{2}}+\bar{\delta}_{\phi}+\hat{\delta}_{\phi}\right)$\tabularnewline
			\hline
			\hline 
						$\delta_{v}=\frac{1}{2}\left(\bar{\delta}_{\phi}+\hat{\delta}_{\phi}\right)$ & 
			$\delta_{v}|_{\text{FJ}}=\frac{1}{2}\left(\delta_{\mu^{2}}-\delta_{\lambda}\right)$ \tabularnewline
			\hline
			$\delta t_h =  -\lambda v^{3}\left(\delta_{\lambda}-\delta_{\mu^{2}} +  \bar{\delta}_{\phi} + \hat{\delta}_{\phi} \right)$
			&
			$\delta t_h|_\mathrm{FJ} =  0$
			\tabularnewline
			\hline 
			$\delta_{M_{h}^{2}}=\frac{3}{2}\left(\delta_{\lambda}+\bar{\delta}_{\phi}+\hat{\delta}_{\phi}\right)-\frac{1}{2}\delta_{\mu^2}$ & $\delta_{M_{H}^{2}}|_{\text{FJ}}=\delta_{\mu^{2}}$\tabularnewline
			\hline 
			$\delta_{M_{W}^{2}}=2\delta_{g_{2}}+\bar{\delta}_{\phi}+\hat{\delta}_{\phi}$ & $\delta_{M_{W}^{2}}|_{\text{FJ}}=2\delta_{g_{2}}+\delta_{\mu^{2}}-\delta_{\lambda}$  \tabularnewline
			\hline 
			$\delta_{M_{Z}^{2}} = 2\frac{g_{2}^{2}\delta_{g_{2}}+g_{1}^{2}\delta_{g_{1}}}{g_{1}^{2}+g_{2}^{2}} 
			+\bar{\delta}_{\phi}+\hat{\delta}_{\phi} $ 
			& $\delta_{M_{Z}^{2}}|_{\text{FJ}}=2\frac{g_{2}^{2}\delta_{g_{2}}+g_{1}^{2}\delta_{g_{1}}}{g_{1}^{2}+g_{2}^{2}} 
			+\delta_{\mu^{2}}-\delta_{\lambda}$  \tabularnewline
			\hline 
			$\delta_{m_{f}}=\delta_{y}+\frac{1}{2}\left(\bar{\delta}_{\phi}+\hat{\delta}_{\phi}\right)$ & $\delta_{m_{f}}|_{\text{FJ}}=\delta_{y}+\frac{1}{2}\left(\delta_{\mu^{2}}-\delta_{\lambda}\right)$ \tabularnewline
			\hline 
			\hline
		\multicolumn{2}{|c|}{	$\delta_e = \delta_e|_\mathrm{FJ} = \frac{1}{g_{1}^{2}+g_{2}^{2}}\left(g_{1}^{2}\delta_{g_{2}}+g_{2}^{2}\delta_{g_{1}}\right)$  }\tabularnewline
			\hline
		\end{tabular}
		\caption{ 
			The mass, VEV and electric charge renormalization constants are expressed in terms of the renormalization constants of gauge couplings $g_{1}$, $g_{2}$ and potential parameters $\lambda$, $\mu$ together with the (background) field renormalization constants in the two tadpole schemes. This is to emphasize the relations of divergence structures between the different renormalization constants.
			$\Delta$ is the ``FJ term'' used to relate counterterms from the usual tadpole scheme to the FJ-scheme (see Sec. \ref{sec:FJ} ).}
		\label{tab:Application-of-an}
	\end{table}
	\egroup
	
\end{widetext}

\section{Relations among different VEV-schemes}\label{sec:relating-schemes}

\subsection{FJ-scheme}\label{sec:FJ}
The FJ-scheme is a procedure of reinstating tadpole contributions in perturbative calculations so that parameter definitions
can be defined in a gauge-independent way.
In the current literature, this procedure is often paraphrased as a shift of the VEV to the correct minimum of the loop-corrected scalar potential \citep{Denner:2016etu, Denner:2019vbn, Krause:2016gkg,Krause:2016oke}. This means that one first shifts the \emph{bare} VEV by the full tadpole contributions, \ie
\begin{align}\label{eq:FJ-shift1}
	&v_\text{bare} = v_\text{bare}|_\text{FJ}  + \Delta v, \quad \Delta v = \frac{T_h}{m^2_h}\,,
\end{align}
where we indicated the shifted VEV by ``FJ". Only then, one adopts multiplicative renormalization constants and inserts $\Delta v$ at each appearance of the VEV in the Lagrangian before using the parameter relations of \eqref{eq:SM-relations}.
Since \eqref{eq:FJ-shift1} constitutes a redefinition of the bare VEV, it no longer can be interpreted in terms of the background field renormalization constant and we introduce an independent $\delta_v|_\text{FJ}$ for its renormalization.
Then the initial bare VEV is related to the renormalized VEV $v$ as: 
\begin{equation}
	v_\text{bare} = v+  v\, \delta_v|_\text{FJ}  + \Delta v. \label{eq:FJ-shift2} 
\end{equation}
The full tadpole counterterm becomes
\begin{equation}
	\delta t_h=  -\lambda v^{3}\left(\delta_{\lambda}-\delta_{\mu^{2}}+2 \delta_v|_\text{FJ} \right) - \Delta v m^2_h. \label{eq:FJ tadpole counterterm} 
\end{equation}
Using the definition of the FJ-VEV-shift, \eqref{eq:FJ-shift1}, we see that the first term of  \eqref{eq:FJ tadpole counterterm} must vanish in order to fulfill the tadpole condition of \eqref{eq:full tadpole condition} which leads to the identification 
\begin{equation}
\delta_v |_{\text{FJ}} = \frac{1}{2} \left(
\delta_{\mu^2} - \delta_\lambda \right). \label{eq:VEV-FJ-ren}
\end{equation}
This shows that in the FJ-scheme, one recovers the tree-level relation between the VEV and the parameters of the scalar potential of \eqref{eq:minimum}, so that the VEV renormalization can be fully expressed in terms of shifts of the potential parameters $\mu^2$ and $\lambda$. 
In this sense, the shift of \eqref{eq:FJ-shift1} can be  paraphrased as shifting the VEV to the \textit{correct one-loop minimum}.
Or, as presented in~\cite{Fleischer:1980ub}, one chooses the \emph{proper VEV}.

Introducing the shift of \eqref{eq:FJ-shift1} into the Lagrangian everywhere 
is effectively equivalent to including regular tadpole contributions in Green's and vertex functions in addition to the 1PI contributions, \ie effectively setting $\delta t_h = 0$ as well as $\Delta = 0$.
Nevertheless, the FJ-scheme has the appeal that renormalized one-point functions vanish exactly with the tadpole condition \eqref{eq:full tadpole condition} being fulfilled, while they would remain formally divergent in the latter case.
Since \eqref{eq:VEV-FJ-ren} does not include $\hat{\delta}_\phi$ in its definition, it has no gauge-dependent UV-divergences in contrast to the PRTS.
This is shown in Tab.~\ref{tab:Application-of-an} where we also see that none of the other parameters carries a $\hat{\delta}_\phi$-dependence in the FJ-scheme.
Here, we see the relationship between the FJ-scheme and the \emph{additional} VEV-renormalization constant of \cite{Sperling:2013eva}, or equivalently $\hat{\delta}_\phi$: 
the FJ-scheme cancels the gauge-dependent divergences which come about due to the breaking of the global gauge symmetry via the $R_\xi$ gauge-fixing.
We will discuss the gauge-dependence of finite parts at the end of the section.

\subsection{FJ-reparameterization}
\eqref{eq:FJ-shift1} together with the tadpole condition \eqref{eq:full tadpole condition} might seem as two renormalization steps.
However, \eqref{eq:FJ-shift1} represents a mere redefinition of a bare parameter.
In the following, we make this redefinition more explicit and thereby give an equivalent way of formulating the FJ-scheme.
This will also simplify the comparison of the different tadpole schemes.

We first generate counterterms via multiplicative renormalization of the parameters $\{v, \mu^2, \lambda, g_1, g_2\}$ and then do a simple zero-insertion of what we call an \emph{FJ-term} $\Delta$ in the bare VEV, \ie
\begin{equation}\label{eq:Delta-insertion}
	v_\mathrm{bare} = v (1 + \delta_v -\Delta + \Delta)
\end{equation}
where $\Delta$ is the dimensionless equivalent of the VEV-shifts used in \eqref{eq:FJ-shift1}, namely
\begin{equation}
	\Delta = \frac{T_h}{v m^2_h}. \label{eq:Delta-definition}
\end{equation}
We then redefine VEV counterterm by
\begin{equation}
	\delta_v|_\text{FJ} = \delta_v - \Delta\,. \label{eq:delta v FJ}
\end{equation}
which gives us back \eqref{eq:FJ-shift2}, and thereby make the relation between the two tadpole schemes explicit.
Similarly, we can identify
\begin{subequations}\label{eq:masses-FJ-vs-PRTS}
	\begin{align}
		\delta_{m_{f}}|_{FJ}=\delta_{m} - \Delta\,,  \label{eq:fermion FJ} \\
		\delta_{m_{V}^{2}}|_{FJ}=\delta_{m_{V}^{2}} - 2\Delta\,, \label{eq:vector FJ}\\
		\delta_{m_{h}^{2}}|_{FJ}=\delta_{m_{h}^{2}} - 3\Delta\,,\label{eq:del higgs} 
	\end{align}
\end{subequations}
and thereby parameterize the change in going from the renormalization constants of the PRTS to  the ones of the FJ scheme.
In deriving the last line of \eqref{eq:del higgs}, care needs to be taken by imposing the tadpole condition only after the shift \eqref{eq:delta v FJ} was inserted.

Diagrammatically, \eqref{eq:Delta-insertion} can be understood as a simple reassignment of the two instances of $\Delta$.
In the case of one-loop corrections to a fermion mass, this means 
	\begin{align}
		&m_{\text{pole}} = m_{\text{ren}}   +  i\Big(
		\;
		\begin{gathered}
			\vspace{-1.5mm}
			\includegraphics{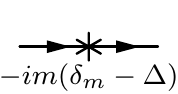}
		\end{gathered}
		\;
		+
		\;
		\begin{gathered}
			\vspace{-1.5mm}
			\includegraphics{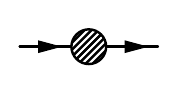}
		\end{gathered}
		\; \nonumber \\ 
		& + 
		\;	
		\begin{gathered}
			\vspace{-1.5mm}
			\includegraphics{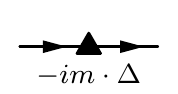}
		\end{gathered}
		\;
		+
		\;
		\begin{gathered}
			\vspace{-5mm}
			\includegraphics{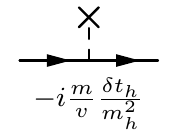}
		\end{gathered}
		\;
		+
		\;
		\begin{gathered}
			\vspace{-5mm}
			\includegraphics{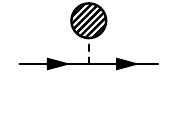}
		\end{gathered}
		\;
		\Big)_{\slashed{p} = m_{\mathrm{ren}}}\,, \label{eq:full-2pt-FJ}
	\end{align} 
where one can identify the FJ mass renormalization constant of \eqref{eq:fermion FJ} in the first diagram, which has no gauge-dependent divergences due to the cancellation of $\hat{\delta}_\phi$ in its definition while the third diagram represents \emph{implicit} tadpole contributions.
Instead, if we choose $\Delta=0$, the first diagram represents a gauge-dependent mass counterterm $\delta_m$ as defined in the PRTS.
In any case, the usual tadpole condition of \eqref{eq:full tadpole condition} provokes the cancellation of the last two diagrams.

Generically, we can write
	\begin{align}
		\begin{gathered}
			\vspace{-1.5mm}
			\includegraphics{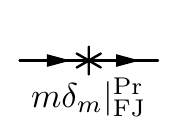}
		\end{gathered}
		= -
		\Big(
		\begin{gathered}
			\vspace{-1.5mm}
			\includegraphics{2pt-blob.pdf}
		\end{gathered}
		+ 
		\begin{gathered}
			\vspace{-1.5mm}
			\includegraphics{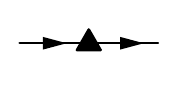}
		\end{gathered}
		\Big)\Big\vert^{\mathrm{Pr}},
		\label{eq:mass-CT-in-FJ}
	\end{align} 
in order to emphasize the inclusion of (implicit) tadpole contributions in the FJ-scheme.
Here, the superscript ``$\mathrm{Pr}$" stands for renormalization prescriptions such as using the \msbar- or OS-scheme, which needs to be specified for discussing the gauge-dependence of finite parts.
Concerning the latter, one finds that the FJ-scheme also leads to gauge independent one-loop mass parameter definitions in the OS-scheme.

There is a simple explanation of why the reintroduction of tadpoles in the FJ-scheme gives a gauge independent result.
As an example, we consider a fermion one-loop pole mass in a bare perturbation theory:
	\begin{equation}
		m_\text{pole}=m_{\text{bare}}   + i\Big(
		\begin{gathered}
			\vspace{-1.5mm}
			\includegraphics{2pt-blob.pdf}
		\end{gathered}
		+
		\begin{gathered}
			\vspace{-1.5mm}
			\includegraphics{2pt-t-blob.pdf}
		\end{gathered}
		\Big)_{\slashed{p} = m_{\mathrm{pole}}}.\label{eq:bare+loop+tad-1}
	\end{equation}
The whole expression \eqref{eq:bare+loop+tad-1} is gauge-independent as
it is a one-loop pole mass \citep{Gambino:1999ai}, while $m_{\text{bare}}$
is gauge-independent by principle. 
Thus the mass shift, i.e. the term in parentheses in \eqref{eq:bare+loop+tad-1}, is gauge independent as well.
This means that the gauge dependence of the tadpole contributions, induced by the Goldstone boson tadpoles of \eqref{eq:Tadpoles-with-Goldstone}, is canceled when added to the 1PI contributions.
With the FJ-scheme implicitly including the tadpole contributions, one realizes that the mass renormalization constants of \eqref{eq:masses-FJ-vs-PRTS} therefore are gauge-independent in terms of their finite parts as well when defined in an OS-scheme.
Note that in the PRTS, where \eqref{eq:full tadpole condition} enforces tadpoles to vanish in \eqref{eq:bare+loop+tad-1} and no VEV-shift is introduced, $m_\text{bare}$ needs to compensate for the gauge-dependence of the 1PI contributions for $m_\mathrm{pole}$ to be gauge-independent (see. \eg \cite{Denner:2016etu}).

\subsection{Numerical comparison \label{sec:direct comparisons} }

In this section, we want to show an explicit determination of the VEV-renormalization constants in the various schemes and give a numerical example for their comparison.

We can use the relations \eqref{eq:SM-relations} to define the VEV-counterterms via the renormalization constants $\delta g_2$ and $\delta m_W^2$ as
\begin{equation}\label{eq:vev-reexpressed}
	\delta v (\{\delta p\}) = \delta v \big(\delta g_2, \delta m_W^2\big) = v \left(\frac{\delta m_W^2}{2 m_W^2} - \frac{\delta g_2}{g_2}\right).
\end{equation}
which  holds for both tadpole schemes.
The $SU(2)$-coupling $g_2$ can in turn be expressed as 
\begin{equation}\label{eq:del-g2-del-e}
	\delta g_2 = \frac{e}{2 s_w} \frac{\delta e + \delta m_W^2 - c_w^2 \, \delta m_Z^2  }{
	m_Z^2 - m_W^2}.
\end{equation}
This step is usually done such that physical quantities known to high accuracy can be used as input parameters.
The charge renormalization constant $\delta e$ can be defined in terms of the $\gamma$-$\gamma$ self energy and the $\gamma$-$Z$ mixing \cite{Denner:2019vbn}.
It does not depend on the choice of a tadpole scheme (because the Higgs field $h$ does not couple to the photon and there is no $\gamma$-$Z$-$h$ vertex) and neither does $\delta g_2$ as defined in \eqref{eq:del-g2-del-e}.
Instead, the difference between the VEV renormalization constants comes about via the definition of $\delta m_W^2$.
As explained in Sec.~\ref{sec:FJ}, the FJ-scheme is equivalent to including tadpoles in counterterm definitions either implicitly as in \eqref{eq:mass-CT-in-FJ}, or explicitly as if they were not renormalized at all.
This means  we can define
	\begin{equation}
		\delta m_W^2\vert_{\mathrm{FJ}} = \mathrm{Re}\Big[ -i\Big(
		\begin{gathered}
			\vspace{-1.5mm}
			\includegraphics{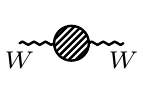}
		\end{gathered}
		+
		\begin{gathered}
			\vspace{-1.5mm}
			\includegraphics{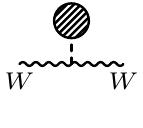}
		\end{gathered}
		\Big)\Big]^{\mathrm{transverse}}_{p^2 = m_W^2}, \label{eq:Mw-FJ}
	\end{equation}
for the on-shell $W$-boson mass counterterm. 
This definition is gauge-independent.
In contrast to that, no tadpole contributions enter \eqref{eq:vev-reexpressed} in the PRTS, leading to the definition
	\begin{equation}
		\delta m_W^2\vert_\mathrm{PRTS} = \mathrm{Re}\Big[ -i\Big(
		\begin{gathered}
			\vspace{-1.5mm}
			\includegraphics{2pt-W.pdf}
		\end{gathered}
		\Big)\Big]^{\mathrm{transverse}}_{p^2 = m_W^2},\label{eq:Mw-PRTS}
	\end{equation}
which is a gauge-dependent quantity. Hence using \eqref{eq:Mw-FJ} or  \eqref{eq:Mw-PRTS} in \eqref{eq:vev-reexpressed} defines the VEV counterterm in the FJ or the PRTS scheme respectively.
A third version of the VEV counterterm is the definition via the Higgs background field renormalization $\delta_{\hat{h}}$ using \eqref{eq:renorm-fields-vev}.
One possible renormalization condition for the latter is
\begin{align}\label{eq:del-hat-h-OS}
	\frac{\partial }{ \partial p^2} \Sigma^{R}_{\hat{h}\hat{h}}|_{p^2=m_h^2} = 0,
\end{align}
where $\Sigma^{R}_{\hat{h}\hat{h}}$ is the one-loop renormalized self-energy of the Higgs background field.
Then, we can define
\begin{equation}\label{eq:delv-BG}
	\delta v|_\text{BG}=\frac{v}{2} \delta_{\hat{h}}|_\mathrm{OS}.
\end{equation}
A definition of the VEV counterterm in terms of field renormalization constants can be expected to be gauge-dependent and this is indeed the case.

Now, using \eqref{eq:vev-reexpressed} together with \eqref{eq:Mw-FJ} for the FJ-scheme and \eqref{eq:Mw-PRTS} the PRTS, 
one can verify by an explicit one-loop calculation that 
\begin{align}\label{eq:FJ-vs-Dominik}
	\delta v\vert_{\mathrm{PRTS}} =
	\delta v\vert_{\mathrm{FJ}} + \Delta v \overset{\infty}{=}  \delta v|_\text{BG}.
\end{align}
This serves as a consistency check of \eqref{eq:delta v FJ} together with \eqref{eq:field and vev}.
Note that the second equal sign only holds for the UV-parts of the counterterms which is to be expected, because there is no simple relation between \eqref{eq:del-hat-h-OS} and the other renormalization conditions in terms of finite parts. 
Nevertheless, we think it is illuminating to see a direct comparison of three rather different approaches to tadpole and VEV renormalization.

Using the numerical input parameters of App. \ref{app:numerics}, the expressions above yield 
\begin{subequations}
	\begin{align}
		\delta v_{\mathrm{fin}}|_\text{PRTS} &= 10.155\,\text{GeV}, \label{eq:finite-PRST-CT} \\
		\delta v_{\mathrm{fin}}|_\text{FJ} &= - 138.457\,\text{GeV},\label{eq:finite-FJ-CT} \\
		\Delta v_{\mathrm{fin}} &= 148.612\,\text{GeV},\label{eq:finite-FJ-shift} \\
		\delta v_\text{fin}|_\text{BG} &= 15.2504 \,\text{GeV}. \label{eq:finite-VEV-CT-ours}
	\end{align}\label{eq:numerical-comparison}
\end{subequations}
With these quantities being renormalization scale dependent and all of them except for \eqref{eq:finite-FJ-CT} being gauge-dependent, one can not put too much of an interpretation to these values.
Nevertheless, it allows us to discuss two interesting aspects.

Firstly, we see that \eqref{eq:finite-FJ-CT}, being defined OS, and \eqref{eq:finite-FJ-shift} show a large cancellation when combined to give \eqref{eq:finite-PRST-CT}, which we tested for a wide range of renormalization scale values.
In theories where not all model parameters can be defined via process-independent physical observables in an OS-scheme, such cancellations can be absent, potentially leading to sizable shifts when $\Delta$ is included in parameter definitions (for the sake of gauge-independence).

Secondly, both the VEV-shift as well as $\delta v_{\mathrm{fin}}|_\text{FJ}$ receive their largest contribution from the heaviest SM-particle, the top-quark.
While in the SM, one could minimize the numerical effect of $\Delta v$ or $\delta v_{\mathrm{fin}}|_\text{FJ}$ by choosing a renormalization scale $\mu_R$ such that the top-quark tadpole vanishes\footnote{To be precise, $\delta v_{\mathrm{fin}}|_\text{FJ}$ vanishes at a scale of $\mu_R^2 \simeq (182\, \mathrm{GeV})^2/\exp(1)$.},
this would not be of help in theories with particles at much higher mass scales.
This is because even if $\mu_R$ is chosen to minimize the effect of the heaviest particle tadpoles (\ie via minimizing the effect of $\ln(\mu_R^2/m_\mathrm{heavy}^2)$ in the resulting $A_0$-functions),
the lighter ones would in turn yield sizable shits due to the large scale difference between $\mu_R$ and $m_\mathrm{light}$.

These aspects are in accordance with the findings of~\cite{Freitas:2002pe} where it was shown that tadpole contributions can be a source of numerical instabilities.
Another example is the multi-Higgs Doublet Standard Model discussed by one of the authors in~\cite{Grimus:2018rte}, where an \msbar-scheme is used and heavy Majorana fermions give large contributions to $\Delta v$, ultimately leading to very large corrections for one-loop neutrino masses. 

In this sense, the FJ-scheme can present a trade-off between gauge-independent quantities and numerical stability in a perturbative calculation.

\subsection{Remarks}
\begin{enumerate}
		
	\item The considerations in Sec.~\ref{sec:FJ} are helpful for adjusting parameter input values in one-loop calculations from the FJ-scheme to ones without tadpole contributions.
	As an example, we mention the \msbar-scheme, where from \eqref{eq:delta v FJ} one finds
	\begin{equation}
		v^{\overline{\text{MS}}}|_\text{FJ} = v^{\overline{\text{MS}}}(1-\Delta_\text{finite}), \label{eq:FJ vs usual vev}
	\end{equation}
	and similar relations for the masses via \eqref{eq:masses-FJ-vs-PRTS}.
	In addition, since the l.h.s.\ of \eqref{eq:FJ vs usual vev} is gauge independent, one can fully account for the gauge-dependencies on the r.h.s.\ via the Goldstone contributions of $\Delta$. 
	
	\item It is rather straightforward to generalize the procedure of Sec.~\ref{sec:FJ} to BSM
	models with altered scalar sectors and we refer to the existing presentations of \cite{Denner:2019vbn,Denner:2016etu}.
	The Two-Higgs Doublet Model is one example.
	Here, a simple way to generalize the FJ procedure is to choose the Higgs basis~\cite{Davidson:2005cw, Haber:2006ue, Haber:2010bw}.
	Then one can straightforwardly use \eqref{eq:masses-FJ-vs-PRTS} with $\Delta$ being identified by \eqref{eq:Delta-definition}.
	The only difference is that $T_h$ and $\delta t_h$ now represent the tadpole contributions and the tadpole counterterms of the extended scalar sector in the Higgs basis.
	Another example is the multi-Higgs doublet model as discussed in~\cite{Grimus:2018rte}, where one can attribute an FJ-term $\Delta_k$ to each doublet $\phi_k$ and define these via the respective tadpole contributions $T_k$.
	
	\item One can think about solving the issues of gauge-dependence, numerical stability and moreover the compatibility of the background-field approach to VEV-renormalization by investigating alternative definitions of $\Delta$.
	One alternative is to define
	\begin{equation}
		\Delta =  \frac{ T_h^{(\xi)}}{v m_h^2},
	\end{equation}
	where $T_h^{(\xi)}$ are only the gauge-dependent parts of the tadpole contributions, \ie the Goldstone boson tadpoles of \eqref{eq:Tadpoles-with-Goldstone}.
	This could be justified by the claim that these contributions come about as an artifact of the global symmetry breaking effect of the $R_\xi$-gauge.
	In this approach, gauge-independence as in the original FJ-scheme would still be guaranteed, because the difference in the choices of $\Delta$ would only lie in gauge-independent contributions.
	Then, large contributions to $\Delta$, \eg from the top quark or even heavier particles in BSM models would be absent and lead to numerically smaller corrections.
	Nevertheless, one could argue that this choice of $\Delta$ seems somewhat arbitrary in the sense that not all tadpole contributions are treated on the same footing and that moreover, the extraction of the  gauge-dependent tadpole contributions might become less obvious at higher loop-orders.
	
	\item
	As discussed in Sec. \ref{sec:Additional ct},  $\hat{\delta}_\phi$ absorbs the gauge-dependent divergences coming from the introduction of the  $R_\xi$-gauge. 
	One could try to mimic the effect of an FJ-VEV shift via an appropriate choice of the finite parts of $\hat{\delta}_\phi$.
	An obvious attempt would be to use an OS-condition for the renormalized two-point function of the physical Higgs field and the  Higgs background field\footnote{This is in contrast to the approach of \cite{Sperling:2013eva}, where only infinite contributions are taken into account.}, \ie
	\begin{equation}
		\frac{\partial }{ \partial p^2} \Sigma_{hh}|_{p^2=m_h^2} = 0, \quad \frac{\partial }{ \partial p^2} \Sigma_{\hat{h}\hat{h}}|_{p^2=m_h^2} =0.\label{eq:OS fields}
	\end{equation}
	This yields
	\begin{align}
		\hat{\delta}_\phi|_\text{OS} &= \frac{1}{2}(\delta_h|_\text{OS}-\delta_{\hat{h}}|_\text{OS}) \nonumber \\
		&=  \frac{1}{4}\xi g_{2}^{2}\frac{1}{\left(4\pi\right)^{2}}\Big[2B_{0}\left(m^2_h,\xi m_{W}^{2},\xi m_{W}^{2}\right) \nonumber \\
		& +\frac{1}{\cos^{2}\theta_{W}}B_{0}\left(m^2_h,\xi m_{Z}^{2},\xi m_{Z}^{2}\right)\Big]\,. \label{eq:delta hat OS}
	\end{align}
	Remarkably, it gives the same functional expression as the unphysical Green's function used to check the divergences in \eqref{eq:gamma q K}, except that the 
	subtraction point is $p^2=m_h^2$ instead of $p^2=0$. 
	This choice of $\hat{\delta}_\phi$ is equivalent to using \eqref{eq:delv-BG}.
	The numerical comparison in \eqref{eq:numerical-comparison} shows no coincidence with the other VEV-renormalizations and therefore indicates that the renormalization conditions used in the PRTS are incompatible with \eqref{eq:delta hat OS}.
	
	In App.~\ref{app:OS gauge dependence}, we show that when promoting \eqref{eq:tadpole gauge condition} to a condition on finite terms as well, so that $\hat{\delta}_\phi$ absorbs the full Goldstone tadpoles, it leads to a gauge dependent charge renormalization condition. This setting is practically equivalent to using $\hat{\delta}_\phi|_\text{OS}$, apart from subtraction point being $p^2=0$ instead of $p^2=m^2_h$. Hence it shows the incompatibility of  $\hat{\delta}_\phi|_\text{OS}$  with the usual charge renormalization condition as discussed in Sec.~\ref{sec:direct comparisons}. 
	
	In general, there seems to be no obvious way to define the finite parts of $\hat{\delta}_\phi$ in order to mimic the effect of the FJ-procedure and it therefore remains purely as a tool for studying divergence structures.
\end{enumerate}

\section{Conclusions}\label{sec:conclusions}
The symmetry breaking effect of the $R_\xi$-gauge fixing leads to the necessity of a renormalization constant in addition to the ones for parameters and fields for the option to render all $n$-point Green's functions finite.
By adapting the findings of \cite{Sperling:2013eva} to the SM,
we showed explicitly how this independent renormalization constant is related to the Higgs background field renormalization and to Goldstone boson tadpoles. 
Effectively this degree of freedom was used in the tadpole condition already in~\cite{Passarino:1978jh}, yet we wanted to emphasize its origin lying in the gauge-fixing as opposed to being a direct consequence of spontaneous symmetry breaking.
We have shown how this degree of freedom leads to gauge dependencies in all the counterterms it enters, such as the mass counterterms in Tab.~\ref{tab:Application-of-an}.

The FJ-scheme \cite{Fleischer:1980ub} manages to avoid these gauge dependencies in parameter and counterterm definitions.
The scheme was originally presented with arguments about 
using the \emph{proper} VEV, while in the recent literature the notion of \emph{true one-loop minimum} \cite{Denner:2019vbn} was employed.
We, however, showed that one can look at this scheme as being simply a set of convenient counterterm redefinitions and in this way provide some independence from the interpretation of these notions.

The global symmetry argument which let us claim that we need only one additional renormalization constant $\hat{\delta}_\phi$ breaks down whenever $\xi_W\neq \xi_Z$.
On the other hand, the FJ-scheme generalizes to any loop-order straightforwardly also when $\xi_W\neq \xi_Z$, even though it implicitly uses the 
degree of freedom of the Higgs background field renormalization.
This hints at the possibility that a single renormalization constant $\hat{\delta}_\phi$ is enough also in this case, yet it is unclear whether there exists a rigorous symmetry argument for that. 
Hence, there is a subtle interplay between the VEV-renormalization, interpreted as the renormalization of the Higgs background field, and the FJ-scheme.
 
We used the SM as a playground to test various aspects of the renormalization in the $R_\xi$-gauge with a special emphasis on tadpole conditions and the connection between different approaches to the subject.
This becomes especially relevant in BSM models with extended scalar sectors, where \eg numerical effects of tadpole contributions and the discussion of gauge-dependencies in mixing angles remain actively discussed \cite{Freitas:2002pe,Denner:2018opp}.
We advocate the use of the FJ prescription for keeping track of gauge dependencies in intermediate expressions and as a useful tool for consistency checks in perturbative calculations.
Nevertheless, we also draw attention to settings where the FJ-scheme can yield large corrections in renormalized quantities, potentially leading to numerical instabilities in non-OS schemes and discuss a possible alternative.

\begin{acknowledgments}

\section*{Acknowledgments}

The authors are grateful to Thomas Gajdosik for discussions and a careful reading of the manuscript. We would also like thank João P.\ Silva and Dominik Stöckinger for encouraging us to carry out this project.
M.L.\ also thanks Marcel Krause and Stefan Liebler for many helpful discussions.

V.D.\ thanks the Lithuanian Academy of Sciences for their support
via project DaFi2019.
M.L.\ is supported partially by the DFG Collaborative Research Center TRR 257 “Particle Physics Phenomenology after the Higgs Discovery”.
\end{acknowledgments}

\appendix

\section{Gauge fixing functions\label{app:Gauge-fixing-functions} }

In the background field formalism, the gauge-fixing functions for the $U\left(1\right)$ and $SU\left(2\right)$
gauge fixing parts respectively, are given by: 
\begin{equation}
F_{B}=\partial^{\mu}B_{\mu}-i\xi g_{1}\left(\hat{\phi}^{\dagger}\frac{Y}{2}\phi-\phi^{\dagger}\frac{Y}{2}\hat{\phi}\right)\,,\label{eq:Gauge fix B}
\end{equation}
\begin{equation}
F_{W}^{i}=\partial^{\mu}W_{\mu}^{i}-i\xi g_{2}\left(\hat{\phi}^{\dagger}\frac{\sigma^{i}}{2}\phi-\phi^{\dagger}\frac{\sigma^{i}}{2}\hat{\phi}\right)\,,\label{eq:Gauge fix W}
\end{equation}
where $\phi$ is the Higgs doublet field and $\hat{\phi}$ is its background field. 
Both fields, $\phi$ and $\hat{\phi}$, transform in the same way
under the global gauge transformation, hence it is easy to show that
\eqref{eq:Gauge fix B} is invariant under the gauge transformation,
while $F_{W}^{i}$ of \eqref{eq:Gauge fix W} transforms as a
vector in the adjoint representation of $SU\left(2\right)$. The anti-ghost
$\bar{c}^{i}$ also transforms as a vector in the adjoint representation,
hence the gauge-fixing term,
\begin{equation}
\mathcal{L}_{GF}=s\left[\bar{c}^{i}\left(F^{i}+\frac{\xi}{2}B^{i}\right)\right]\,,\label{eq:gauge fix lagrangian}
\end{equation}
is invariant. The mass eigenstate gauge-fixing functions are recovered by: 
\begin{equation}
F_{A}=F_{W}^{3}s_{W}+F_{B}c_{W}\,,\label{eq:FA-1}
\end{equation}
\begin{equation}
F_{Z}=F_{W}^{3}c_{W}-F_{B}s_{W}\,,\label{eq:FZ-1}
\end{equation}
\begin{equation}
F_{W^{\pm}}=\frac{1}{\sqrt{2}}\left(F^{1}\mp iF^{2}\right)\,,\label{eq:FW-1}
\end{equation}
where $s_W$ and $c_W$ are sine and cosine of the Weinberg angle, respectively. 
Taking the limit
\begin{equation}
	\hat{\phi}\to\left(\begin{array}{c}
		0\\
		\frac{1}{\sqrt{2}}v
	\end{array}
\right)
\end{equation}
 in \eqref{eq:Gauge fix B} and \eqref{eq:Gauge fix W} and inserting them into \eqref{eq:FA-1}, \eqref{eq:FZ-1}, and \eqref{eq:FW-1}, we recover the usual $R_{\xi}$-gauge fixing functions: 
\begin{align}
F_{A} &=\partial^{\mu}A_{\mu}, \nonumber \\
F_{Z} &=\partial^{\mu}Z_{\mu}-\xi m_{Z}G_{Z},\nonumber \\
F_{W}^{\pm} &=\partial^{\mu}W_{\mu}^{\pm}\mp i\xi m_{W}G_{W}^{\pm}.
\end{align}

\section{Calculation using BRS-sources}
\label{app:Calculation-using-BRS}

Following \cite{Sperling:2013eva} we find the divergence structure of
$\hat{\delta}_{\phi}$ directly from the unphysical Green's functions that include
BRST sources.  The Higgs doublet of the SM is decomposed into quantum and background field: 
\begin{equation}
\Phi=\phi+\hat{\phi}\,.
\end{equation}
The fields $\Phi$ and $\hat{\phi}$ correspond to $\phi^{\text{eff}}$
and $\hat{\phi}+\hat{v}$ of \cite{Sperling:2013eva} respectively.
The BRST transformation of the background field is postulated to be
in a contractible pair with another background field $\hat{q}$:
\begin{equation}
s\hat{\phi}=\hat{q}_{\phi}\,,\,\,s\hat{q}_{\phi}=0\,,\label{eq:BRS back-1}
\end{equation}
so neither of $\hat{\phi}$, nor $\hat{q}$ contribute to the BRST cohomology \cite{Barnich:2000zw}. In other words, they do not contribute
to the physical spectrum of the theory. The BRST transformation
for the field $\Phi$ is
\begin{equation}
s\Phi=g_{2}i\frac{\sigma^{k}}{2}\left(\phi+\hat{\phi}\right)c_{k}+\frac{i}{2}g_{1}\left(\phi+\hat{\phi}\right)c_{B}\,.\label{eq:BRS full-1}
\end{equation}
From \eqref{eq:BRS back-1} and \eqref{eq:BRS full-1} we get
\begin{equation}
s\phi=\left(g_{2}i\frac{\sigma^{k}}{2}\left(\phi+\hat{\phi}\right)c_{k}+\frac{i}{2}g_{1}\left(\phi+\hat{\phi}\right)c_{B}-\hat{q}_{\phi}\right)\,,\label{eq:BRS quant-1}
\end{equation}
where $\sigma$ are Pauli matrices and $c_k$ and $c_B$ are the ghost fields of the $SU(2)$ and $U(1)$ gauge group respectively. 
Finally, we include the BRST source $K_{\phi}$ in the Lagrangian
\begin{equation}
\mathcal{L}_{K}=K_{\phi}^{\dagger}s\phi+s\phi^{\dagger}K_{\phi}\,. \label{eq:BRS lagr}
\end{equation}

\begin{figure}
		\begin{subequations}
		\begin{align}
		&\begin{gathered}
			\vspace{-1.5mm}
			\includegraphics{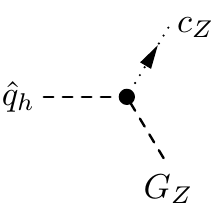}
		\end{gathered}
		= -i \xi \frac{g_2}{2 c_W}, &
		\begin{gathered}
			\vspace{-1.5mm}
			\includegraphics{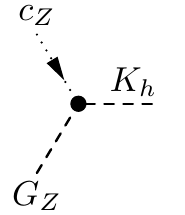}
		\end{gathered}
		= i \frac{g_2}{2 c_W}, \nonumber \\[25pt]
		&\begin{gathered}
			\vspace{-1.5mm}
			\includegraphics{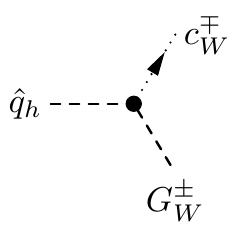}
		\end{gathered}
		= \pm \xi \frac{g_2}{2}, &
		\begin{gathered}
			\vspace{-1.5mm}
			\includegraphics{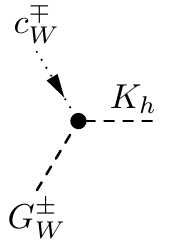}
		\end{gathered}
		= \pm \frac{g_2}{2}, \nonumber \\[25pt]
		&\begin{gathered}
			\vspace{-1.5mm}
			\includegraphics{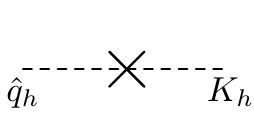}
		\end{gathered}
		= -i \hat{\delta}_\phi. \nonumber
		\end{align} 
	\end{subequations}
	\caption{Feynman rules for calculating $\hat{\delta}_{\phi}$.\label{fig:Feynman-rules-for delz}}
\end{figure}
\begin{figure}
			\begin{align}
			\begin{gathered}
				\vspace{-1.5mm}
				\includegraphics{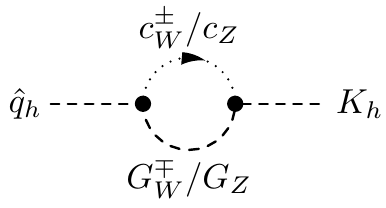}
			\end{gathered} \nonumber
			\end{align}
	\caption{One-loop diagram for calculating $\hat{\delta}_{\phi}$. \label{fig:BRS loop}}
\end{figure}

The renormalization transformations of field and background field from \eqref{eq:renormalization+bf} are
\begin{equation}
\phi\to \sqrt{ \bar{Z}_{\phi} / \hat{Z}_{\phi}  }\, \phi\,,\,\,\hat{\phi}\to \sqrt{ \bar{Z}_{\phi} \hat{Z}_{\phi} }\,\hat{\phi}\,, \label{eq:field and bfield transf}
\end{equation}
thus the introduced ``technical'' background field $\hat{q}$ transform as:
\begin{equation}
s\hat{\phi} \to s\sqrt{\bar{Z}_{\phi}\hat{Z}_{\phi}}\,\hat{\phi}= \sqrt{\bar{Z}_{\phi}\hat{Z}_{\phi}}\, \hat{q} \quad
\Rightarrow \quad \hat{q}\to \sqrt{\bar{Z}_{\phi}\hat{Z}_{\phi}}\,\hat{q}\,.
\end{equation}
BRST sources transform as the inverse renormalization transformation of the corresponding field. Then the relation
\begin{equation}
\frac{\delta\Gamma}{\delta K_{\phi}}=\left\langle s\phi\right\rangle\,,
\end{equation}
where $\Gamma$ is the effective vertex functional, is unchanged after the renormalization transformation. From \eqref{eq:field and bfield transf} we get that the transformation for the BRST source of the Higgs doublet quantum field is: 
\begin{equation}
K_{\phi}\to \sqrt{ \hat{Z}_{\phi} / \bar{Z}_{\phi} }\, K_{\phi}\,.\label{eq:K transformation-1-1}
\end{equation}
Including all these renormalization transformations into \eqref{eq:BRS lagr} we get:
\begin{align}
\mathcal{L}_{K} & =K_{\phi}^{\dagger}\left(g_{2}i\frac{\sigma^{k}}{2}\left(\phi+\hat{Z}_{\phi}\hat{\phi}\right)c_{k}+\frac{i}{2}g_{1}\left(\phi+\hat{Z}_{\phi}\hat{\phi}\right)c_{B}\right)\label{eq:brs of fi-1-1-1-1}\\
& -\hat{Z}_{\phi}K_{\phi}^{\dagger}\hat{q}_{\phi}+h.c.\nonumber 
\end{align}
The last term gives the counterterm for an unphysical
Green's function that includes only $\hat{Z}_{\phi}$. 
We will look at the Green's function $\Gamma_{\hat{q}_{h}K_{h}}$, where 
$h$ is the Higgs field component of the doublet. The counterterm of this 
Green's function is shown in the last diagram of Fig.~\ref{fig:Feynman-rules-for delz}.
To calculate this Green's function at one-loop, one only needs
the Feynman rules for interactions between $\hat{q}$, $K_{\phi}$ and $c$, which can be read out from \eqref{eq:brs of fi-1-1-1-1}
and \eqref{eq:gauge fix lagrangian}, and are shown in Fig.~\ref{fig:Feynman-rules-for delz}.
The loop diagram that we will need to calculate is shown in Fig.~\ref{fig:BRS loop}.
The result of the sum of Fig.~\ref{fig:BRS loop} and the last diagram of Fig.~\ref{fig:Feynman-rules-for delz}
is:\begin{widetext}
	\begin{align}
	i\hat{\Gamma}_{\hat{q}_{h}K_{h}}^{\left[1\right]}= & -i\hat{\delta}_{\phi}+i\frac{1}{4}\xi g_{2}^{2}\frac{1}{\left(4\pi\right)^{2}}\left[2B_{0}\left(0,\xi m_{W}^{2},\xi m_{W}^{2}\right)+\frac{1}{\cos^{2}\theta_{W}}B_{0}\left(0,\xi m_{Z}^{2},\xi m_{Z}^{2}\right)\right]\label{eq:gamma q K}
	\end{align}
	or, using $A_{0}\left(m^{2}\right)=m^{2}\left(1+B_{0}\left(0,m^{2},m^{2}\right)\right)\,$,
	$m_{W}=\frac{g_{2}v}{2}$ and $m_{W}=\cos\theta_{W}m_{Z}$:
	\begin{align}\label{eq:BRST-del-phi-hat}
	\hat{\Gamma}_{\hat{q}_{h}K_{h}}^{\left[1\right]}= & -\hat{\delta}_{\phi}+\frac{1}{\left(4\pi\right)^{2}v^{2}}\left\{2\left[A_{0}(m_{W}^{2}\xi)-\xi m_{W}^{2}\right] + \left[A_{0}(m_{Z}^{2}\xi)-\xi m_{Z}^{2}\right]\right\}\,.
	\end{align}
\end{widetext}The finiteness of this two-point function fixes the divergences of $\hat{\delta}_{\phi}$.
Moreover, \eqref{eq:BRST-del-phi-hat} immediately shows that its divergences coincide with the ones of the Goldstone boson tadpoles \eqref{eq:Tadpoles-with-Goldstone},
which means that $\hat{\delta}_\phi$ indeed absorbs all the gauge-dependent divergences
in the tadpole condition \eqref{eq:full tadpole condition}.

\section{Gauge dependence in the background-field-modified OS-scheme \label{app:OS gauge dependence}}

In this section, we present the consequence of promoting \eqref{eq:tadpole gauge condition} to 
a renormalization condition on finite parts, namely:  
\begin{equation}
T_{h}^{(\xi)}-\lambda v^{3}\hat{\delta}_{\phi} =0.\label{eq:tadpole gauge condition finite}
\end{equation}
In principle, this is a valid renormalization condition, which allows to absorb 
all the tadpole finite gauge dependencies into $\hat{\delta}_{\phi}$. One can
verify that the FJ-term $\Delta$ is
\begin{equation}
\Delta= \frac{1}{2} \hat{\delta}_{\phi} + \text{gauge-independent}, \label{eq:delta with hat} 
\end{equation}
when \eqref{eq:tadpole gauge condition finite} holds. This allows for a more direct 
interpretation of an FJ term in the sense of a background field renormalization also in finite 
parts, leading to a gauge-independent OS mass renormalization constant.  Also, the form of \eqref{eq:delta with hat}
 gives a possibility to modify the FJ procedure 
to include only the gauge-dependent term, namely  $\hat{\delta}_{\phi}$.
 However, 
we will show that this choice, together with the OS conditions on the two point functions,
leads to a gauge-dependent charge renormalization constant and cannot be used with the usual charge renormalization 
condition presented in \eg \cite{Denner:1991kt, Boehm:1333727, Denner:2019vbn}.

To show this, we first need to get the gauge-dependent part of 
$\bar{\delta}_\phi$ in this scheme. For that we look at
the gauge-dependent part of the Higgs self energy, which can be written as:
\begin{align}
\Sigma_{\xi}&=\frac{1}{\left(4\pi v\right)^{2}}\left(p^{2}-m_{h}^{2}\right)\left(f_{Z}+2f_{W}\right)\nonumber  \\ 
&+m_{h}^{2}\frac{1}{\left(4\pi v\right)^{2}}\frac{3}{2}\left(A_{0}\left(m_{Z}^{2}\xi_{Z}\right)+2A_{0}\left(m_{W}^{2}\xi_{W}\right)\right) \,, \label{eq:Higgs self gauge}
\end{align}
\begin{align}
f_{V}=A_{0}\left(m_{V}^{2}\xi\right)-\frac{1}{2}\left(p^{2}+m_{h}^{2}\right)B_{0}\left(p^{2},m_{V}^{2}\xi,m_{V}^{2}\xi\right)\,.\label{eq:Higgs self gauge add}
\end{align}
The gauge dependent part of the renormalized one-loop self energy function is 
\begin{align}
\Sigma_{\xi}^{R} & =\delta_{h}|_{\xi}p^{2}-\left(\delta_{m_{h}^{2}}+\delta_{h}\right)|_{\xi} \, m_{h}^{2}+\Sigma_{\xi}\,, \label{eq:Higgs self ren}
\end{align}
where we write $|_\xi$ to denote the gauge-dependent terms of the renormalization constants. 
The OS conditions give:
\begin{equation}
\frac{\partial}{\partial p^2}\Sigma^R |_{p^2=m^2_H}= 0\,,\quad \Sigma^R |_{p^2=m^2_H}= 0\,. \label{eq:OS conditions}
\end{equation} 
Inserting \eqref{eq:Higgs self gauge}, \eqref{eq:Higgs self gauge add}, \eqref{eq:Higgs self ren} into OS conditions \eqref{eq:OS conditions} to check the gauge-dependent parts, we get the gauge dependencies of mass and field renormalization constants of the OS-scheme (\ie in the tadpole scheme of ref. \cite{Denner:1991kt} ):
\begin{align}
&\delta_{m_{h}^{2}}|_{\xi}=\frac{1}{\left(4\pi v\right)^{2}}\frac{3}{2}\left(A_{0}\left(m_{Z}^{2}\xi_{Z}\right)+2A_{0}\left(m_{W}^{2}\xi_{W}\right)\right)\,, \label{eq:mass gauge dependence} \\
 &\delta_{h}|_{\xi}  =  
 - \frac{1}{(4\pi v)^{2}} \Big\{\Big[ A_{0}(m_{Z}^{2}\xi) + 2 A_{0}(m_{W}^{2}\xi) \Big] \nonumber \\
  &
  -
    m_{h}^{2}  \Big[ B_{0}(p^{2},m_{Z}^{2}\xi,m_{Z}^{2}\xi) + 2 B_{0}(p^{2},m_{W}^{2}\xi,m_{W}^{2}\xi) \Big]\Big\}\,.
\end{align}

By using the expression for $\Delta$ from \eqref{eq:delta with hat} and $\hat{\delta}_\phi$, fixed by \eqref{eq:tadpole gauge condition finite}, we see that the FJ-OS mass counterterm, as defined in \eqref{eq:del higgs}, is truly gauge
independent: 
\begin{equation}
\left(\delta_{m_{h}^{2}}|_{FJ}\right)|_{\xi}=\delta_{m_{h}^{2}}|_{\xi}-3\Delta_{\xi}=\delta_{m_{h}^{2}}|_{\xi}-\frac{3}{2}\hat{\delta}_{\phi}=0\,.\label{eq:FJ gauge independence}
\end{equation}
From \eqref{eq:field and vev} we can get the field renormalization part $\bar{\delta}_{\phi}$ that does not have gauge-dependent divergences. However, it turns out that in the OS, the finite part of $\bar{\delta}_{\phi}$ is gauge-dependent:
\begin{align}
\bar{\delta}_{\phi} |_\xi &= \delta_h |_\xi+ \hat{\delta}_\phi  \nonumber \\
 &=\frac{1}{(4\pi v)^{2}} m_{h}^{2} \Big[ B_{0}(p^{2},m_{Z}^{2}\xi,m_{Z}^{2}\xi) \nonumber \\
 &\qquad \qquad \qquad + 2 B_{0}(p^{2},m_{W}^{2}\xi,m_{W}^{2}\xi) \Big]\,. \label{eq:gauge dependence of phi}
\end{align}
Note that the divergences in this term are gauge-independent as it should be.

Now using \eqref{eq:tadpole gauge condition finite} together with \eqref{eq:Delta-definition} we see that we must have:
\begin{equation}
\delta_{\lambda}|_{\xi} + \bar{\delta}_{\phi}|_{\xi} - \delta_{\mu^2}|_{\xi} = 0. \label{eq:gauge dependence consistency}
\end{equation}
From the fact that the FJ mass renormalization constant of the Higgs coincides with $\delta_{\mu^2}$ (see. Tab \ref{tab:Application-of-an}) and is gauge-independent, the gauge dependence of $\delta_\lambda$ is: 
\begin{equation}
\delta_{\lambda}|_{\xi} =- \bar{\delta}_{\phi}|_{\xi}. \label{eq:gauge dependence of lam}
\end{equation}

From Tab. \ref{tab:Application-of-an}, we see that $\delta_\lambda$ enters the definition of the FJ mass renormalization constants, which are gauge-independent. Hence from the gauge independence of $\delta_{M^2_W}|_\text{FJ}$, $\delta_{M^2_Z}|_\text{FJ}$ and $\delta_{m_f}|_\text{FJ}$ we get 
\begin{align}
&0 =2\delta_{g_{1,2}}|_\xi-\delta_{\lambda}|_\xi = 2\delta_{g_{1,2}}|_\xi + \bar{\delta}_{\phi}|_\xi \,, \\
&0=\delta_{y}|_\xi-\frac{1}{2}\delta_{\lambda}|_\xi = \delta_{y}|_\xi + \frac{1}{2}\bar{\delta}_{\phi}|_\xi \,.
\end{align}
which leads to
\begin{equation}
\frac{1}{g_{1}^{2}+g_{2}^{2}}\left(g_{1}^{2}\delta_{g_{2}}|_{\xi}+g_{2}^{2}\delta_{g_{1}}|_{\xi}\right) +\bar{\delta}_{\phi}|_\xi= \delta_e|_\xi +\bar{\delta}_{\phi}|_\xi =0 \,. \label{eq:gauge dependence of del e}
\end{equation}
From \eqref{eq:gauge dependence of del e}, we see that the charge renormalization constant $\delta_e$ is gauge dependent, because it needs to cancel the gauge-dependence of $\bar{\delta}_{\phi}|_\xi$ given in \eqref{eq:gauge dependence of phi}.
Note that this is solely because we enforced \eqref{eq:tadpole gauge condition finite}. 
Thus, we have two tadpole conditions, pole and residue conditions for $W$, $Z$ and Higgs two point function, which are in total 8 conditions, hence fully determines $g_{1,2}$, $\delta_\lambda$, $\delta_{\mu^2}$, $\bar{\delta}_\phi$, $\hat{\delta}_\phi$ and field renormalization constants of Z and W bosons. This means that there is no freedom left to impose
a charge renormalization condition as in \eg \cite{Denner:1991kt, Boehm:1333727, Denner:2019vbn}, which would
give a gauge-independent charge renormalization constant otherwise. Nevertheless, it is interesting to see that the gauge
independent definition of the charge renormalization constant is possible also in this ``scheme'' by absorbing $\bar{\delta}_\phi$ 
into their definitions, as suggested by \eqref{eq:gauge dependence of del e}. Hence in principle it is possible to 
use \eqref{eq:tadpole gauge condition finite} instead of the usual charge renormalization condition and even define a gauge-independent 
charge renormalization constant. Yet to understand all the consequences of this unconventional choice a more thorough study is needed which is beyond the scope of this work.

\section{Explicit divergences \label{app:explicit divs}}

We used the SM-model file from FeynArts together with FormCalc to get the explicit expressions for the divergences.
The divergences of 1, 2, 3 and 4 point function of the Higgs in the
SM respectively are:
\begin{align}
&\Gamma_{h}^{\text{UV}}=v^{3}\left(A+G+S\right)\,,\\
&\Gamma_{hh}^{\text{UV}}=p^{2}\left(\frac{1}{v^{2}}B+\frac{1}{2m_{h}^{2}}G\right)+v^{2}\left(3A+5S+G\right)\,, \\
&\Gamma_{hhh}^{\text{UV}}=6v\left(A+2S-G\right)\,,\\
&\Gamma_{hhhh}^{\text{UV}}=6\left(A+2S-2G\right)\,,\label{eq:4pt divs}
\end{align}
where we abbreviated 
\begin{align}
A= &-\frac{1}{\left(4\pi v^{2}\right)^{2}} \Big[ 4 \big(\sum_{f=e,\mu,\tau}m_{f}^{4}+3\sum_{q=u,d,s,c,t,b}m_{q}^{4} \big) \\
 & -3 (2m_{W}^{4}+m_{Z}^{4} )\ \Big]\,,\\
B= &\frac{1}{ (4\pi v^{2} )} \Big[2 \big(\sum_{f=e,\mu,\tau}m_{f}^{2}+3\sum_{q=u,d,s,c,t,b}m_{q}^{2} \big) \\
&  -3 (2m_{W}^{2}+m_{Z}^{2} ) \Big]\,,\\
S= &\frac{3}{2}\frac{m_{h}^{4}}{ (4\pi v^{2} )^{2}}\,, \quad
G= \frac{1}{2}\frac{m_{h}^{2}}{ (4\pi v^{2} )^{2}} (2m_{W}^{2}\xi+m_{Z}^{2}\xi )\,,
\end{align}
and omitted a global factor of $\frac{2}{4-D}$. The divergences in $A$ and $B$ come
from loop diagrams with ghosts, vectors and fermions. Note that neither
$A$ nor $B$ are gauge-dependent, since ghost and vector boson gauge
dependencies cancel exactly. The divergences abbreviated as $S$ come
from diagrams with the Higgs boson loop contribution. Finally, the
only gauge-dependent UV divergent term, $G$, corresponds to divergences
of the Goldstone boson loop and vanishes in case of $\xi\to0$. Note
that we wrote all these functions in terms of 4 abbreviated constants,
thus we need 4 independent conditions and 4 degrees of freedom to
uniquely fix them. In case of $\xi\to0$, the number is reduced to
only three. The four degrees of freedom are, $\mu$, $\lambda$, $\bar{\delta}_{\phi}$
and $\hat{\delta}_{\phi}$. They appear in the counterterms of one-, two-, three-
and four-point functions of the Higgs respectively:
\begin{align}
&\delta\Gamma_{h}=-\frac{1}{2}m_{h}^{2}v\left(\delta_{\lambda}-\delta_{\mu^{2}}+\bar{\delta}_{\phi}+\hat{\delta}_{\phi}\right)\,, \\
&\delta\Gamma_{hh}=-\frac{1}{2}m_{h}^{2}\left(3\delta_{\lambda}-\delta_{\mu^{2}}+5\bar{\delta}_{\phi}+\hat{\delta}_{\phi}\right)+p^{2}\left(\bar{\delta}_{\phi}-\hat{\delta}_{\phi}\right)\,,\\
&\delta\Gamma_{hhh}=-3\frac{m_{h}^{2}}{v}\left(\delta_{\lambda}+2\bar{\delta}_{\phi}-\hat{\delta}_{\phi}\right)\,,\\
&\delta\Gamma_{hhhh}=-3\frac{m_{h}^{2}}{v^{2}}\left(\delta_{\lambda}+2\bar{\delta}_{\phi}-2\hat{\delta}_{\phi}\right)\,.\label{eq:4pt counter}
\end{align}
To make sure that the renormalized n-point functions are finite, we solve four equations: 
\begin{equation}
\delta\Gamma_{h}^{\text{UV}}+\Gamma_{h}^{\text{UV}}=0\,,
\end{equation}
\begin{equation}
\delta\Gamma_{hh}^{\text{UV}}+\Gamma_{hh}^{\text{UV}}=0\,,
\end{equation}
\begin{equation}
\frac{\partial}{\partial p^{2}}\left(\delta\Gamma_{hh}^{\text{UV}}+\Gamma_{hh}^{\text{UV}}\right)=0\,,
\end{equation}
\begin{equation}
\delta\Gamma_{hhh}^{\text{UV}}+\Gamma_{hhh}^{\text{UV}}=0\,.
\end{equation}
They give us 
\begin{align}
 & \delta_{\mu^{2}}=B+\frac{1}{\lambda}S\,,\,\,\delta_{\lambda}=2\left(B+\frac{1}{\lambda}S\right)+\frac{1}{\lambda}A\,,\nonumber \\
 & \hat{\delta}_{\phi}=\frac{1}{\lambda}G\,,\,\,\bar{\delta}_{\phi}=-B\,,\quad\lambda=\frac{m_{h}^{2}}{2v^{2}}\,.\label{eq:divs of constants}
\end{align}
Inserting these expressions into \eqref{eq:4pt counter} we automatically
get
\begin{equation}
\delta\Gamma_{hhhh}^{\text{UV}}+\Gamma_{hhhh}^{\text{UV}}=0\,,
\end{equation}
where $\Gamma_{hhhh}^{\text{UV}}$ is given in \eqref{eq:4pt divs}. 

The VEV-counterterms, presented in Sec.~\ref{sec:direct comparisons} have divergences, which can be read out
from Tab.~\ref{tab:Application-of-an}, inserting the expressions for the renormalization constants,
shown in \eqref{eq:divs of constants}:
\begin{align}
&\delta v=\frac{1}{2}v\left(\hat{\delta}_{\phi}+\bar{\delta}_{\phi}\right)=\frac{1}{2}v\left(\frac{1}{\lambda}G-B\right)\,,\\
&\delta v|_{\text{FJ}}=\frac{1}{2}v\left(\delta_{\mu^{2}}-\delta_{\lambda}\right)=-\frac{1}{2}v\left(B+\frac{1}{\lambda}S+\frac{1}{\lambda}A\,\,\right),\\
&\delta v|_{\text{FJ}}  =\delta v+\Delta v\nonumber \\
&\Rightarrow  \Delta v=-\frac{1}{2}v\left(\frac{1}{\lambda}S+\frac{1}{\lambda}A+\frac{1}{\lambda}G\right)=-\frac{1}{m_{h}^{2}}\Gamma_{h}^{\text{UV}}\,.
\end{align}

\section{Input values \label{app:numerics}}

Here we present the numerical input values used in Eqs.~(\ref{eq:finite-PRST-CT} - \ref{eq:finite-VEV-CT-ours}) together with the software package LoopTools \cite{Hahn_1999} at its standard renormalization scale of $\mu_R = 1\,\mathrm{GeV}$.

\begin{align}\label{eq:numerical-input-SM}
& \xi=1, \; m_W =  80.398\,\text{GeV},\;m_Z =  91.1876\,\text{GeV},
\;\nonumber \\ 
& m_h =  125.09\,\text{GeV},\;  v =  246.221\,\text{GeV} \nonumber \\
& m_e =  0.000510999\,\text{GeV},\; m_\mu =  0.105658\,\text{GeV},\; \nonumber\\
& m_\tau =  1.77684\,\text{GeV},\;\nonumber  \\
& m_u =  0.19\,\text{GeV},\; m_c=  1.4\,\text{GeV},\; m_t =  172.500\,\text{GeV},\; \nonumber \\
& m_d =  0.19\,\text{GeV},\; m_s =  0.19\,\text{GeV},\; m_b =  4.75\,\text{GeV},\; \nonumber \\
& e =    0.308147. \nonumber
\end{align}

\bibliographystyle{apsrev4-1.bst}
\bibliography{vev-bib.bib} 

\end{document}